\documentclass[aps,prb,twocolumn,showpacs]{revtex4}
\usepackage{amsmath}
\usepackage{dcolumn}
\usepackage{amssymb}
\usepackage{amsfonts}
\usepackage{graphicx}
\usepackage[T1]{fontenc}
\usepackage{hyperref}

\begin{document}

\title{Bosonic Fractional Quantum Hall States in Rotating Optical Lattices: \\Projective Symmetry Group Analysis}
\author{T. \DJ uri\'c and A. Lazarides}
\affiliation{Max Planck Institute for the Physics of Complex Systems,
N\"othnitzer Str. 38, 01187 Dresden, Germany}
\date{\today}

\begin{abstract}

  We study incompressible ground states of bosons in a two-dimensional rotating square optical lattice. The system can be described by the Bose-Hubbard model in an effective uniform magnetic field present due to the lattice rotation. To study ground states of the system, we map it to a frustrated spin model, followed by Schwinger boson mean field theory and projective symmetry group analysis. Using symmetry analysis we identify bosonic fractional quantum Hall states, predicted for bosonic atoms in rotating optical lattices, with possible stable gapped spin liquid states within the Schwinger boson formalism. In particular, we find that previously found fractional quantum Hall states induced by the lattice potential, and with no counterpart in the continuum [G. M\"oller, and N. R. Cooper, Phys. Rev. Lett. \textbf{103}, 105303 (2009)], correspond to ``$\pi$ flux'' spin liquid states of the frustrated spin model.

\end{abstract}

\pacs{67.85.-d, 73.43.-f, 75.10.Kt, 75.10.Jm}
\maketitle
\section{Introduction}

\label{sec:intro}

Ultracold atoms in optical lattices have provided new opportunities to experimentally realise and study a wide range of complex quantum models under controllable conditions.\cite{Lewnstein, JakschZoller, Greiner} Many studies of atomic systems in artificial gauge potentials were motivated by the possibility to realize strongly correlated fractional quantum Hall (FQH) states.\cite{Cooper,Bhat, SorensenDemlerLukin, WilkinGunn, CooperWilkinGunn,
  RezayiReadCooper, PalmerJaksch, Hafezi, Palmer, MollerCooper, KapitMueller,
  UmucalilarMueller} On the other hand, quantum Hall effects
\cite{Klitzing, Tsui1, Tsui2} have attracted much attention since
their discovery, and are still a very active field of study. In
addition to being intrinsically interesting systems, they are also
being investigated from the point of view of topological quantum computing:  excitations in some FQH systems exhibit fractional, anyonic statistics and can potentially perform fault-tolerant quantum computation. \cite{Kitaev, DasSarma}

An artificial gauge field for neutral atoms can be created, for example, by
rotating the lattice.\cite{Cooper, Bhat, CooperWilkinGunn, Parades, Polini,
  Tung} The rotation of the system is equivalent, in the corotating frame, to
the introduction of an effective magnetic field proportional to the rotation
frequency.\cite{Cooper,Bhat} However, the system enters the FQH regime at a
very high rotation frequency or a low atomic density which are hard to achieve
experimentally. This has motivated development of many alternative schemes
\cite{JakschZoller, SorensenDemlerLukin, Mueller, Furtado, Pachos, PachosRico,
  Kay, Juzeliunas,Osterloh, Ruseckas, Cooper2, CooperDalibard} to introduce a vector potential to a system of neutral atoms, for example, using Raman assisted hopping combined with lattice acceleration or an inhomogeneous static electric field,\cite{JakschZoller, Mueller} or using oscillating quadrupole fields.\cite{SorensenDemlerLukin}

Prior to studies of FQH states in optical lattices, such states have
been predicted for rotating Bose-Einstein condensates in harmonic
magnetic traps.\cite{WilkinGunn, CooperWilkinGunn} The rotation
(effective magnetic field) introduces vortices into the condensate and
eventually, at sufficiently high rotation frequency, leads to
formation of an Abrikosov vortex lattice.  \cite{Cooper,
  WilliamsHolland} When the number of vortices becomes comparable to the number of bosons the system can enter into a FQH state. However, since the interactions among the bosonic atoms in the magnetic traps are weak, the FQH ground state is separated from the excited states by a small energy gap. On the other hand, in optical lattices interaction energies among the atoms are much larger due to confinement of atoms in a smaller volume. This leads to more robust FQH states separated from the excited states by much larger energy gaps than for the atoms in harmonic traps. In addition, the presence of the lattice leads to new interesting physical effects. In particular, new FQH states with no counterpart in the continuum limit \cite{MollerCooper} were predicted for systems of bosons in the presence of the lattice potential and effective magnetic field.

In the presence of a tight-binding lattice and a uniform magnetic
field the single-particle energy levels change from simple Landau
levels in the absence of the lattice into the Hofstadter butterfly
spectrum.\cite{Hofstadter, UmucalilarOktel} The energy spectrum
depends on the magnetic flux $\alpha$ per lattice plaquette (measured
in units of the flux quantum $\phi_0=2\pi\hbar/Q$, where $Q$ is the
effective charge of the particle). 

For rational $\alpha=p/q$ (with $p$ and $q$ coprime) the spectrum
splits into $q$ bands and each state is $q$ fold
degenerate. When the single-particle energies are plotted against
the flux $\alpha$ the resulting Hofstadter butterfly has a fractal
structure. For $\alpha\ll 1$ the low-lying bands become the Landau
levels of the continuum. For bosonic atoms with repulsive
interactions, in the presence of a lattice and a uniform effective
magnetic field with $\alpha\ll 1$ one thus expects to find the same
states as in a continuum bosonic quantum Hall system. Some of the
states that appear in the continuum are, for example, bosonic
$\nu=1/2$ Laughlin state and $\nu=1,3/2,...$ Read-Rezayi states
\cite{Laughlin, ReadRezayi, Regnault} with $\nu=n/\alpha$ and $n$
being the average particle density per lattice site. It has been shown
that at sufficiently low particle density the lattice has a negligible
effect on the nature of the continuum $\nu=1/2$
state.\cite{SorensenDemlerLukin, Hafezi} However, in the presence of
the lattice additional novel FQH states with no counterpart in the
continuum limit have also been predicted.\cite{MollerCooper}

In the present paper we study incompressible ground states of
interacting bosons in a two dimensional rotating square optical
lattice. Bosonic atoms in a deep optical lattice can be described by a Bose-Hubbard model. An effective vector potential, due to the lattice rotation, introduces an Aharonov-Bohm phase for the bosons hopping from site to site and the complex tunelling couplings appear in the Hubbard Hamiltonian. In the regime of strong on-site interaction the Bose-Hubbard Hamiltonian can be mapped to an effective frustrated spin model. To study the spin model, we use Schwinger boson mean field theory \cite{ArovasAuerbach, Chandra, BurkovMacDonald,
  Sachdev} and projective symmetry group (PSG) analysis.\cite{WangVishwanath, Wang,
  Wen1, Wen2, ZhouWen}

Schwinger boson mean-field theory allows description of both ordered
and disordered phases, that is, the resulting mean field Hamiltonian
does not have any preferred direction in spin space. Magnetic ordering
is identified as Bose-condensation of Schwinger bosons. For such condensed phases the energy excitation spectrum is gapless, while a gapped spectrum corresponds to incompressible spin liquid phases.  In this work, we focus on the possible noncondensed ground states of the system, which do not break any symmetries. For such states, PSG anlysis can be used to determine different mean field ans\"atze that reflect all the physical symmetries of the system and to classify possible spin liquid states.

Unlike conventional states which can be distinguished by patterns of broken symmetry, different spin liquid phases that are completely symmetric can be distinguished by considering how the symmetries are realized in those phases. Such symmetric spin liquid phases can be classified using PSG analysis. The main idea of the PSG analysis of the mean field states in the Schwinger boson theory is that the mean field ansatz preserves all the symmetries of the spin Hamiltonian when it is invariant under transformations that are combined physical symmetry and $U(1)$ gauge group transformations. In other words, symmetries of the spin model can be preserved in the mean field state in some indirect way and invariance of the mean field ansatz under all symmetry transformations is a sufficient, however not necessary condition. The set of all transformations that leave a mean field ansatz invariant is called
the PSG.

Different allowed sets of combined transformations can be found by considering algebraic relations among symmetry group elements. Those different allowed sets of combined transformations then characterize physically distinct symmetric spin states. The bosonic FQH states of the system can then be identified with different spin liquid states of the effective spin model. In particular, our results show that lattice induced FQH states of the system
correspond to ``$\pi$ flux'' spin liquid states of the effective spin
model.

The paper is organized as follows. In Sec. \ref{sec:model} we
introduce the Bose-Hubbard model for the system and outline the
mapping to the quantum spin model. In Sec. \ref{sec:SchwingerBosonMFT}
we describe Schwinger boson mean field theory for the effective
quantum spin model. The magnetic symmetry group operators that
correspond to the symmetries of the system are defined in Sec.
\ref{sec:MSG}. In Sec. \ref{sec:PSG} we find possible noncondensed
spin liquid ground states of the system using PSG analysis. Relation
between the spin liquid states found in Sec. \ref{sec:PSG} and bosonic
FQH states of the system is investigated in Sec. \ref{sec:FQH}. In the final section, we
draw our conclusions.

\section{Effective frustrated spin Hamiltonian}
\label{sec:model}
We consider the system of bosonic atoms in a two-dimensional optical
lattice and in the presence of an effective vector potential introduced, for
example, through lattice rotation. In the limit of weak tunneling $t$ between
wells within the lattice, compared to the level spacing in each well, the
system can be described by a single-band Bose-Hubbard model on a square lattice with a complex hopping matrix element:
\begin{equation}\label{eq:BH}
H_{\rm Hubbard}=H^{(0)}+V 
\end{equation}
with
\begin{eqnarray}\label{eq:BoseHubbard}
H^{(0)}&=&\frac{U}{2}\sum_i a^{\dagger}_i a_i(a^{\dagger}_ia_i -1) - \mu\sum_i a^{\dagger}_ia_i,\nonumber\\
V&=&-t\sum_{\langle ij\rangle}\left(e^{i\phi_{ij}}a_j^{\dagger}a_i+e^{-i\phi_{ij}}a_i^{\dagger}a_j\right),
\end{eqnarray}
where $a_i$ and $a_i^{\dagger}$ are bosonic field operators on site $i$, $U$
is the repulsive energy of two atoms in one well, $\mu$
is the chemical potential and $\langle ij\rangle$ denotes nearest-neighbor
sites $i$ and $j$. The complex tunneling couplings appear in the Hubbard
Hamiltonian due to the presence of the effective vector potential
$\vec{A}$ that introduces an Aharonov-Bohm phase for the
boson hopping from site to site. When an atom moves from a lattice site at
$\vec{r}_i$ to a neighbouring site at $\vec{r}_j$, it will gain an
Aharonov-Bohm phase:
\begin{equation}\label{eq:AharonovBohmPhase}
\phi_{ij}=\int_{\vec{r}_i}^{\vec{r}_j}\vec{A}\cdot d\vec{r}. 
\end{equation}
For a rotating lattice $\vec{A}=m\vec{\Omega}\times\vec{r}/\hbar$, where
$\vec{\Omega}$ is the rotation frequency and $m$ is the mass of the atom. We
consider the case of the uniform magnetic field
$\vec{B}=\vec{\nabla}\times\vec{A}=B\hat{z}$. The vector potential introduces
frustration in the atomic motion if the phase twists around each plaquette add
p to $2\pi\alpha$ for some noninteger $\alpha$. The frustration parameter
$\alpha$ is defined as the flux per plaquette in units of $2\pi$:
\begin{equation}
\alpha=\frac{1}{2\pi}\int\vec{B}\cdot
d\vec{S}_{\rm plaq}=\frac{1}{2\pi}\sum_{\rm plaq}\phi_{ij}, 
\end{equation}
where the integration is over the surface of a lattice plaquette and the sum
is performed anticlockwise over the edges of the square plaquette. Due to the
periodicity under $\alpha\rightarrow\alpha+1$ we can restrict the values of
$\alpha$ to $0\leq\alpha< 1$. The frustration is maximal at $\alpha=1/2$. We choose the Landau gauge $\vec{A}=B\left(0,x,0\right)$ so that the
Aharonov-Bohm phase $\phi_{ij}$ is zero on all horizontal bonds of the
lattice, $\phi_{\left(x,y\right)\left(x\pm 1,y\right)}=0$, and
$\phi_{\left(x,y\right)\left(x, y\pm 1\right)}=\pm2\pi\alpha x$.\\
We further focus on the limit of weak tunneling compared to the repulsive
energy $U$ for two atoms in one well, $t\ll U$, where strong interparticle
repulsion can lead to strongly correlated ground states. In such a limit the
site occupation can be restricted to zero and one boson and the Bose-Hubbard
Hamiltonian (\ref{eq:BH}) can be mapped onto a spin-$1/2$ frustrated $XY$
model. The two $S_z$ states of the pseudospin at a lattice site $i$
correspond to whether a lattice contains a boson or not. The strongly
correlated phases of the system that we study in the present paper will
correspond to spin-liquid phases of the effective frustrated quantum spin
model.\\
The mapping of the Hubbard Hamiltonian (\ref{eq:BH}) in the limit of $t\ll U$ to
an effective spin Hamiltonian is possible because hard-core boson operators have
the same commutation relations as spin-$1/2$ operators. The operators on
different sites commute and operators on the same site anticommute. The spin
raising and lowering operators correspond to the creation and annihilation
operators of hard-core bosons and the motion of the atoms translates to
pseudospin exchange. The effective spin Hamiltonian is:
\begin{equation}\label{eq:easyplanemagnet}
H_{eff}= -\frac{J}{2}\sum_{\langle ij\rangle}
\left(e^{i\phi_{ij}}S_i^+S_j^{-}+e^{-i\phi_{ij}}S_j^+S_i^{-}\right)-h\sum_i\hat{S}_i^z
\end{equation}
where $J=2t$,
$S_{i}^{\pm}=S^x_i \pm iS^y_i$ are spin-$1/2$
operators, and $h=\mu$ represents an effective Zeeman field. Note that
this is a ferromagnet in the absence of frustration ($\phi_{ij} = 0$). Also,
it can be easily verified that the Hamiltonian (\ref{eq:easyplanemagnet}) has
local gauge invariance. If we change the gauge,
$\vec{A}\rightarrow\vec{A}+\vec{\nabla}\chi$, then the Hamiltonian stays
unchanged if the boson and spin operators pick up a phase change,
$\phi_{ij}\rightarrow e^{i\left(\chi_j-\chi_i\right)}\phi_{ij}$,
$a_i\rightarrow e^{i\chi_i}a_i$ and $S_i^{-}\rightarrow
e^{i\chi_i}S_i^{-}$. The change of gauge corresponds to a spin rotation of
$\chi_i$ in the $xy$ plane.\\
In the following sections we study spin liquid phases of the frustrated
quantum spin model (\ref{eq:easyplanemagnet}) using Schwinger boson mean field
theory and PSG analysis. As demonstrated in Sec. \ref{sec:FQH}, those spin
liquid phases correspond to FQH states of the system. 

\section{Schwinger boson mean field theory}
\label{sec:SchwingerBosonMFT}
Schwinger boson mean field theory (SBMFT) was first proposed by Arovas and
Auerbach \cite{ArovasAuerbach} for SU(2) invariant spin models and was further
applied to frustrated antiferromagnets \cite{Chandra} and 
generalized to anisotropic spin models.\cite{BurkovMacDonald} Within the SBMFT
spin--$1/2$ operators are represented in terms of the Schwinger boson
operators
 $a_{i\uparrow}$ and $a_{i\downarrow}$ as follows:
\begin{eqnarray}\label{eq:SchwingerRepresentation}
S_i^z&=&\frac{1}{2}\left(a_{i\uparrow}^{\dagger}a_{i\uparrow}-a_{i\downarrow}^{\dagger}a_{i\downarrow}\right)\;,\\
S_i^{+}&=&a_{i\uparrow}^{\dagger}a_{i\downarrow},\;
S_i^{-}=a_{i\downarrow}^{\dagger}a_{i\uparrow}\;. \nonumber
\end{eqnarray}
In addition, the Schwinger bosons satisfy the local constraint 
\begin{equation}\label{eq:Constraint}
a_{i\uparrow}^{\dagger}a_{i\uparrow}+a_{i\downarrow}^{\dagger}a_{i\downarrow}=1
\end{equation}
at every lattice site. 

To derive a mean field theory that can describe both ordered and disordered
phases, i.e. a mean field Hamiltonian that does not have any preferred
direction in spin space, the first step is to rewrite a spin Hamiltonian
in terms of bond operators which are bilinear forms in Schwinger boson
operators from nearest-neighbor sites. Introduction of one complex
Hubbard-Stratonovich field at each bond then provides a natural mean field
decoupling scheme. There are usually many possible bond operators that one can
define. However, the only relevant ones are the bond operators that are invariant
under the transformations of the symmetry group of the Hamiltonian. The spin
Hamiltonian (\ref{eq:easyplanemagnet}) is invariant under spin
rotations around the $z$-direction and has $U(1)$ symmetry. In that case one
can define four different bond operators that are invariant under symmetry
operations of the Hamiltonian:
\begin{eqnarray}\label{eq:BondOperators}
F_{ij}&=&e^{i\phi_{ij}/2}a_{i\uparrow}^{\dagger}a_{j\uparrow}+e^{-i\phi_{ij}/2}a_{i\downarrow}^{\dagger}a_{j\downarrow}\;,\\
A_{ij}&=&e^{-i\phi_{ij}/2}a_{i\uparrow}a_{j\downarrow}-e^{i\phi_{ij}/2}a_{i\downarrow}a_{j\uparrow}\;, \nonumber\\ 
X_{ij}&=&e^{-i\phi_{ij}/2}a_{i\uparrow}a_{j\downarrow}+e^{i\phi_{ij}/2}a_{i\downarrow}a_{j\uparrow} ,\nonumber\\ 
Z_{ij}&=&e^{i\phi_{ij}/2}a_{i\uparrow}^{\dagger}a_{j\uparrow}-e^{-i\phi_{ij}/2}a_{i\downarrow}^{\dagger}a_{j\downarrow}\;. \nonumber
\end{eqnarray}
It can be easily verified that defined bond operators commute with the
operator for the total magnetization in the $z$-direction,
$\hat{M}=\sum_iS_i^z$, and are thus $U(1)$ symmetric. These bond operators are analogous to the bond operators defined for the
quantum $XXZ$ model \cite{BurkovMacDonald} with additional phase factors which
reflect the presence of an effective vector potential.\\
Nonzero expectation values for these bond operators signal the presence of
various short-range correlations, that is, indicate the presence of short-range
order. $F_{ij}$ and $A_{ij}$ are analogous to $SU(2)$ ferromagnetic and antiferromagnetic bond
operators, respectively. The bond operators $X_{ij}$
and $Z_{ij}$ represent $XY$ and easy axis correlations,
respectively. Depending on the expected dominant correlations in the ground
state of a given Hamiltonian, the Hamiltonian is further rewritten in terms
of the products of appropriate bond operators. 

For the Hamiltonian
(\ref{eq:easyplanemagnet}) at $h<h_C(\alpha)$, where $h_C(\alpha)$
is a critical value of the effective Zeeman field above which the system is in
the Mott insulating phase, the classical ground state has a finite total magnetization
in the $XY$ plane \cite{DuricLee} and the operators appropriate to describe
the correlations in the ground state are $F_{ij}$ and $X_{ij}$. The
Hamiltonian (\ref{eq:easyplanemagnet}) can then be rewritten in the following form:
\begin{eqnarray}\label{eq:Hamiltonian}
  H&=&\frac{J}{4}\sum_{\langle ij\rangle}n_{i}\cdot n_j+\frac{Jz}{8}\sum_i n_i\\
  &-&\frac{J}{4}\sum_{\langle
  ij\rangle}\left(F_{ij}^{\dagger}F_{ij}+X_{ij}^{\dagger}X_{ij}\right)-h\sum_i S_i^z ,\nonumber               
\end{eqnarray}
where $n_i=n_{i\uparrow}+n_{i\downarrow}=2S$ is the number of bosons at each
lattice site and is constant, and $z=4$ is the number of nearest
neighbors. This representation can then be generalized to $N$ Schwinger boson
flavors \cite{ArovasAuerbach,Sachdev} and the functional integral
representation of the partition function can be systematically expanded in
powers of $1/N$. The mean field Hamiltonian corresponds to a saddle point
solution of the large $N$ action after Hubbard Stratonovich
transformation. Since here we are interested only in the mean field solution,
we do not use the large $N$ language. We assume that the bond operators have nonzero expectation values:
\begin{eqnarray}\label{eq:BondOperatorsMF}
\langle F_{ij}\rangle&=&f_{ij},\\
\langle X_{ij}\rangle&=&x_{ij}, \nonumber
\end{eqnarray}
and perform a Hartree--Fock decomposition of the Hamiltonian (\ref{eq:easyplanemagnet}):
\begin{equation}\label{eq:HartreeFock}
O_{ij}^{\dagger}O_{ij}\rightarrow o_{ij}O_{ij}^{\dagger}+ o_{ij}^{*}O_{ij}-|o_{ij}|^{2},
\end{equation}
where $O_{ij}\in \{F_{ij},X_{ij}\}$. The mean field Hamiltonian is then:
\begin{eqnarray}\label{eq:Hamiltonian_MF}
H_{MF}&=&\frac{J}{4}\sum_{\langle
  ij\rangle}\left(|f_{ij}|^2+|x_{ij}|^2\right)+\frac{J}{4}\sum_{\langle
  ij\rangle}n_{i}\cdot n_j+\frac{Jz}{8}\sum_i n_i \nonumber\\
&-&\frac{J}{4}\sum_{\langle
  ij\rangle} \left(f_{ij}\cdot F_{ij}^{\dagger}+f_{ij}^*\cdot
  F_{ij}+x_{ij}\cdot X_{ij}^{\dagger}+x_{ij}^*\cdot
  X_{ij}\right)\nonumber\\
&-&\frac{h}{2}\sum_i\left(n_{i\uparrow}-n_{i\downarrow}\right) +\sum_i\lambda_i\left(n_{i\uparrow}+n_{i\downarrow}-\kappa\right).             
\end{eqnarray}
Within the mean field theory the constraint on the Schwinger boson occupation
number at site $i$ is imposed on average:
\begin{eqnarray}\label{eq:ConstarintAverage}
\langle n_{i\uparrow}+n_{i\downarrow}\rangle=2S=\kappa,
\end{eqnarray}
by introducing a chemical potential $\lambda_i$ and the average boson number
$\kappa$ can also be taken as a parameter.\cite{Sachdev} Mean field theory
parameters $f_{ij}$ and $x_{ij}$ are in general complex numbers called the
mean field ansatz and can be found by solving self-consistent equations
$\langle F_{ij}\rangle_{MF}=f_{ij}$ and $\langle
X_{ij}\rangle_{MF}=x_{ij}$.

Within this method,  magnetic ordering in the $XY$ plane is identified as Bose condensation of the Schwinger bosons. In the presence of the magnetic ordering the energy excitation spectrum of Schwinger bosons is gapless, while a gapped excitation spectrum corresponds to incompressible spin liquid phases. Further we will focus on the possible noncondensed ground states of the system which are expected to be states without any spontaneously broken symmetry. As explained in one of the following sections, for such states we can use projective symmetry group analysis (PSG) to determine different mean field ans\"atze that reflect all the physical symmetries of the system, and to classify possible spin liquid states that can appear in the phase diagram of the system.

\section{Magnetic symmetry group}
\label{sec:MSG}
Before proceeding to the PSG analysis we define a group of operators that
correspond to the symmetries of the system and which commute with the
Hamiltonian (\ref{eq:easyplanemagnet}). The presence of an effective
vector potential reduces the physical symmetries of the lattice and the
Hamiltonian (\ref{eq:easyplanemagnet}) does not commute with the
standard lattice spatial symmetry operations. However, it is still
possible to define operators that correspond to physical symmetries of
the system. As will be described below, such operators obey the same
multiplication relations as the original lattice spatial symmetry
operators apart from additional phase factors and commute with the
Hamiltonian (\ref{eq:easyplanemagnet}). Generalized translation,
rotation and reflection transformation operators in the presence of a
uniform magnetic field can be specified and from such operators a full
group can be constructed which is known as the magnetic symmetry
group.\cite{Wen2,Brown,Zak,Powell}

\begin{figure}[thb]
\includegraphics[width=1\columnwidth]{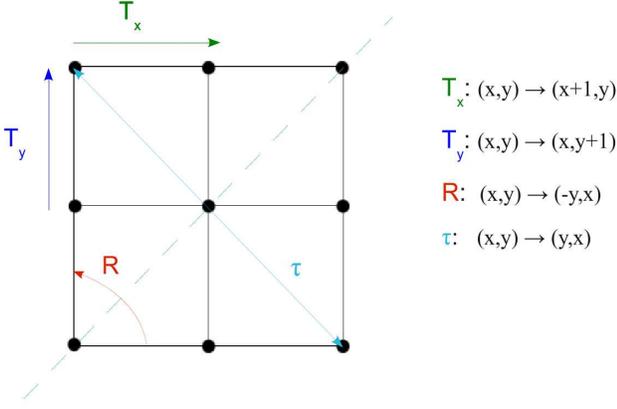}
\caption{(Color online) Symmetry operations (see text).}
\label{symmetry}
\end{figure}

We define elementary translation ($T_x$, $T_y$), rotation ($R$) and reflection
($\tau $) operators, illustrated in Fig. \ref{symmetry}, in terms of their commutation relations with the spin
operators $S_i^{\pm}$. Since we have chosen the Landau gauge, the Hamiltonian (\ref{eq:easyplanemagnet}) commutes with the translation operator in $y$ direction and the operator $T_y$ can be defined by
\begin{equation}\label{eq:Ty}
T_yS_{(x,y)}^{+} =S_{(x,y+1)}^{+}T_y .
\end{equation}
It can be easily verified that $\left[H_{eff},T_y\right]_{-}=0$. However, the operator of a pure translation in $x$ direction does not commute with the Hamiltonian  (\ref{eq:easyplanemagnet}) due to the presence of $x$ dependent phase factors, and we therefore define a generalized unitary translation operator in $x$ direction as follows:
\begin{equation}\label{eq:Tx}
T_xS_{(x,y)}^{+} = e^{-i2\pi\alpha y}S_{(x+1,y)}^{+}T_x ,
\end{equation}
where the phase factors in the previous equation are chosen to ensure
$\left[H_{eff},T_x\right]_{-}=0$. For $\alpha=p/q$ the Hamiltonian has
periodicity $q$ in $x$ direction and $T_x^{q}S_{(x,y)}^{+} =
S_{(x+1,y)}^{+}T_x^{q}$. Similarly one can define a unitary rotation operator giving rotation by $\pi/2$ counterclockwise around the origin:
\begin{equation}\label{eq:R}
R S_{(x,y)}^{+} = e^{i2\pi\alpha xy}S_{(-y,x)}^{+}R ,
\end{equation}
where the phase factor is again chosen to obtain  $\left[H_{eff},R \right]_{-}=0$. Also, an antiunitary operator $\tilde{\tau}$ that corresponds to reflection in the line $y=x$ can be defined,
\begin{equation}\label{eq:tau1}
\tilde{\tau} S_{(x,y)}^{+} = e^{-i2\pi\alpha xy}S_{(y,x)}^{+}\tilde{\tau} ,
\end{equation}
which in combination with the antiunitary complex conjugation operator, $C$, forms a unitary operator and is a symmetry of the Hamiltonian (\ref{eq:easyplanemagnet}):
\begin{equation}\label{eq:tau2}
\tau \lambda S_{(x,y)}^{+} = \lambda^{*} e^{-i2\pi\alpha xy}S_{(y,x)}^{+}\tau ,
\end{equation}
where $\tau=C\tilde{\tau}$, $\lambda$ is a complex number, and
$\left[H_{eff},\tau \right]_{-}=0$. The reflection operator itself is
not a symmetry of the Hamiltonian since it reverses chirality while
the magnetic field explicitly breaks chirality and time
reversal symmetry. However, after combination with complex conjugation
the appropriate sense of circulation is restored and operator $\tau$
denotes a symmetry of the Hamiltonian.

The magnetic symmetry group operators defined above have the same multiplication relations as ordinary spatial group operators apart from additional phase factors:
\begin{eqnarray}\label{eq:MultiplicationRelations}
T_xT_y&=&e^{i2\pi\alpha}T_yT_x ,\\
T_xR&=&RT_y^{-1}\;,\;RT_x=T_yR\;,\;R^{4}=1,\nonumber\\
\tau T_y &=& T_x\tau \;,\; \tau T_x = T_y \tau , \nonumber\\
R\tau R\tau &=& 1 \;,\; \tau\tau=1 . \nonumber
\end{eqnarray}
We also note that using the Schwinger boson representation of the spin operators (\ref{eq:SchwingerRepresentation}), the magnetic symmetry group operators can also be defined in terms of their commutation relations with Schwinger bosons as follows:
\begin{eqnarray}\label{eq:CommutationRelationsSB}
T_ya_{(x,y)\sigma}&=&a_{(x,y+1)\sigma}T_y , \\
T_xa_{(x,y)\sigma}&=& e^{-i2\pi\alpha_\sigma y}a_{(x+1,y)\sigma}T_x , \nonumber \\
R a_{(x,y)\sigma}&=& e^{i2\pi\alpha_\sigma xy}a_{(-y,x)\sigma}R , \nonumber \\
\tau \lambda a_{(x,y)\sigma}&=& \lambda^{*} e^{-i2\pi\alpha_\sigma xy}a_{(y,x)\sigma}\tau , \nonumber
\end{eqnarray}
where $\lambda$ is a complex number, $\sigma \in
\left\{\uparrow,\downarrow\right\}$, $\alpha_{\downarrow}=\alpha/2$,
$\alpha_{\uparrow}=-\alpha/2$ and
$\alpha=\alpha_{\downarrow}-\alpha_{\uparrow}$, which is a definition
convenient for PSG analysis. The equations
(\ref{eq:MultiplicationRelations}) reflect the structure of the
magnetic space group and will play a crucial role in determining possible mean field ans\"atze using PSG analysis as described in the following section. 

\section{Projective symmetry group analysis}
\label{sec:PSG}
In this section we discuss possible spin liquid phases of the Hamiltonian
(\ref{eq:Hamiltonian_MF}). Spin liquid states are spin disordered states that
typically appear in frustrated spin models.\cite{WangVishwanath, Wang} To
study and classify possible spin liquid phases we use a symmetry based
analysis, PSG analysis, originally introduced by Wen and
collaborators.\cite{Wen1, Wen2, ZhouWen} Within this theory we show that FQH
states induced by a lattice potential,\cite{MollerCooper} and with no
counterpart in the continuum, correspond to ``$\pi$ flux'' spin liquid states
of the effective frustrated spin model. To the best of our knowledge
the connection between the spin liquid phases of the effective spin
Hamiltonian and FQH states for bosonic atoms in rotating optical
lattices has not been previously investigated. 

As we are interested in the phases of the system that do not break any symmetries, we require that the Schwinger boson mean field theory reflects all the underlying microscopic symmetries of the spin model (\ref{eq:easyplanemagnet}). This will lead to symmetric spin liquid states. The symmetry transformations include magnetic symmetry group transformations and $U(1)$ spin rotation symmetry. Using PSG analysis one can then construct all symmetry allowed mean field ans\"atze. It is important, however, to note that the microscopic symmetries of the spin model can be preserved in the mean field state in some indirect way. As first noted by Wen and collaborators \cite{Wen1, Wen2, ZhouWen} in the context of fermionic mean field theory of spin liquids, the mean field theory ansatz should be invariant under transformations that are combined physical symmetry and gauge group transformations. This idea was then generalized to Schwinger boson mean field theory for spin liquid states by Wang and Vishwanath.\cite{WangVishwanath, Wang} The transformations that leave the mean field ansatz invariant then form a group called projective symmetry group (PSG). To find PSG transformations we first note that (spin independent) local $U(1)$ gauge transformations of Schwinger bosons 
\begin{equation}\label{eq:GaugeTransformation}
a_{i\sigma}\rightarrow e^{i\phi_i}a_{i\sigma} ,
\end{equation}
leave the Hamiltonian (\ref{eq:easyplanemagnet}) and all physical observables unchanged. However, the mean field ansatz is not invariant with respect to transformations (\ref{eq:GaugeTransformation}) and transforms as follows:
\begin{eqnarray}\label{eq:GaugeTransformation2}
F_{ij}&=&e^{i\left(\phi_j-\phi_i\right)}F_{ij} , \\
X_{ij}&=&e^{i\left(\phi_i+\phi_j\right)}X_{ij} . \nonumber
\end{eqnarray}
Since the physical spin state should be gauge invariant we see that that two
different mean field ans\"atze that are related by a gauge transformation
should, after implementing constraint (\ref{eq:ConstarintAverage}) (that is,
after projection to the physical spin space) lead to the same spin state. In
other words, different mean field ans\"atze may correspond to the same
physical state and the underlying symmetries may not be explicitly present in
the mean field ansatz. Presence of the local $U(1)$ gauge transformations then
explains why invariance of the mean field ansatz with respect to all
microscopic symmetries is not a necessary condition to obtain symmetric spin
liquid states. If the transformed (for example lattice translated) mean field
ansatz is equivalent to $U(1)$ gauge transformed form of the ansatz the same
spin state will be obtained after projection. Accordingly, invariance
with respect to combined physical symmetry and gauge transformations
is a sufficient condition to obtain symmetric states.

In addition to PSG transformations that are related to physical
symmetries, one finds that there are also some elements of PSG that
are pure local $U(1)$ gauge transformations. These transformations are
not the result of a physical symmetry and can be associated with the
emergent gauge group that describes obtained spin liquid phase. Such
transformations form a subgroup of PSG which is called invariant gauge
group (IGG). The microscopic theory of the spin liquid states of the
system involves Schwinger boson spinons which are coupled to the
emergent gauge field.\cite{ArovasAuerbach, Wen3, ReadSachdev} When the
spinon spectrum is gapped and the emergent gauge field takes discrete
values \cite{ReadSachdev, MoessnerSondhiFradkin} ($Z_2$ spin liquid),
or emergent gauge theory is $U(1)$ gauge theory with a nontrivial
topological term \cite{LevinWen}, the mean field solution is stable
and is a good starting point for identifying possible spin liquid states.

To examine the stability of a mean field solution one has to consider the fluctuations in $F_{ij}$, $X_{ij}$ and $\lambda_i$ that descibe the collective excitations above the mean field state. $F_{ij}$ and $X_{ij}$ have the amplitude and phase fluctuations. The amplitude fluctuations have a finite energy gap and are not important in discussing the stability and low energy properties of the possible spin liquid phases. The phase fluctuations of the mean field ansatz are gapped in the presence of a Chern-Simons term or spinon-pair condensation (Anderson-Higgs mechanism). Here the latter is achieved if both $F_{ij}$ and $X_{ij}$ are nonzero and in that case the emergent gauge theory is $Z_2$ (Ising) gauge theory. If $F_{ij}$ or $X_{ij}$ is zero the emergent gauge theory is $U(1)$ gauge theory. However, a Chern-Simons term present in the emergent gauge theory due to an effective
uniform magnetic field gives the $U(1)$ gauge boson a nonzero gap. Gapped gauge bosons can mediate only short range interactions between spinons and stable spin liquid states always contain spinons with only short ranged interactions between spinons.\\
Further we will assume $Z_2$ spin liquid states, that is, we will initially assume that both $F_{ij}$ and $X_{ij}$ are nonzero. It is then clear that the only two elements of the IGG are the identity operation $1$ and the IGG generator $-1$. In other words, the only two transformations that leave the mean field ansatz invariant and are pure local $U(1)$ transformations (\ref{eq:GaugeTransformation}) are $b_i\rightarrow e^{\phi_i}b_i$ with $\phi_i=0$ or $\phi_i=\pi$. The spin $U(1)$ rotation symmetry is already realized by considering mean field ansatz, $X_{ij}$ and $F_{ij}$, of the form (\ref{eq:BondOperators}) which is explicitly invariant under spin rotation around $z$ direction. The operations that we further need to consider are magnetic symmetry group operations.

\begin{figure}[thb]
 \includegraphics[width=0.8\columnwidth]{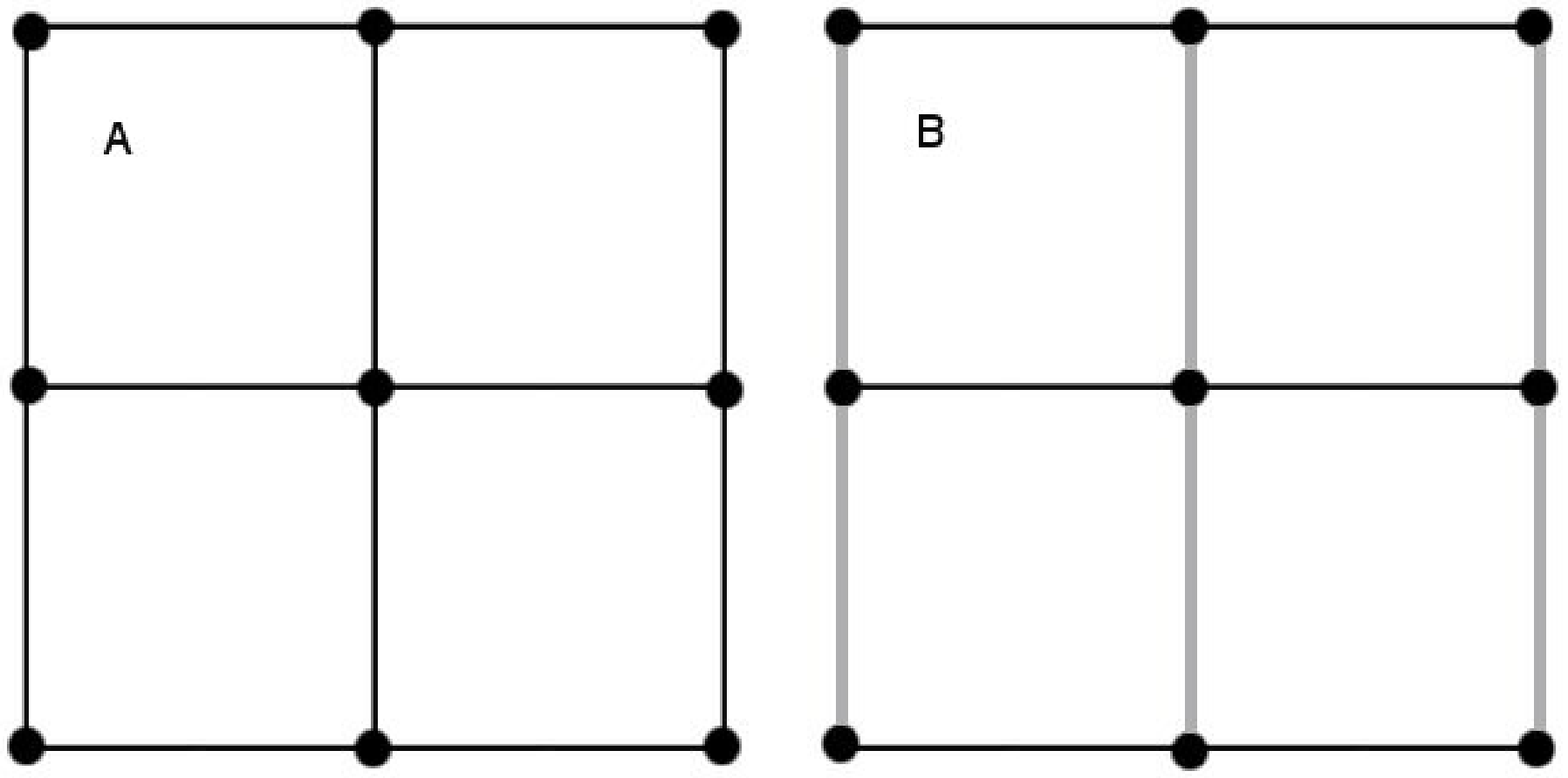}
  \includegraphics[width=0.8\columnwidth]{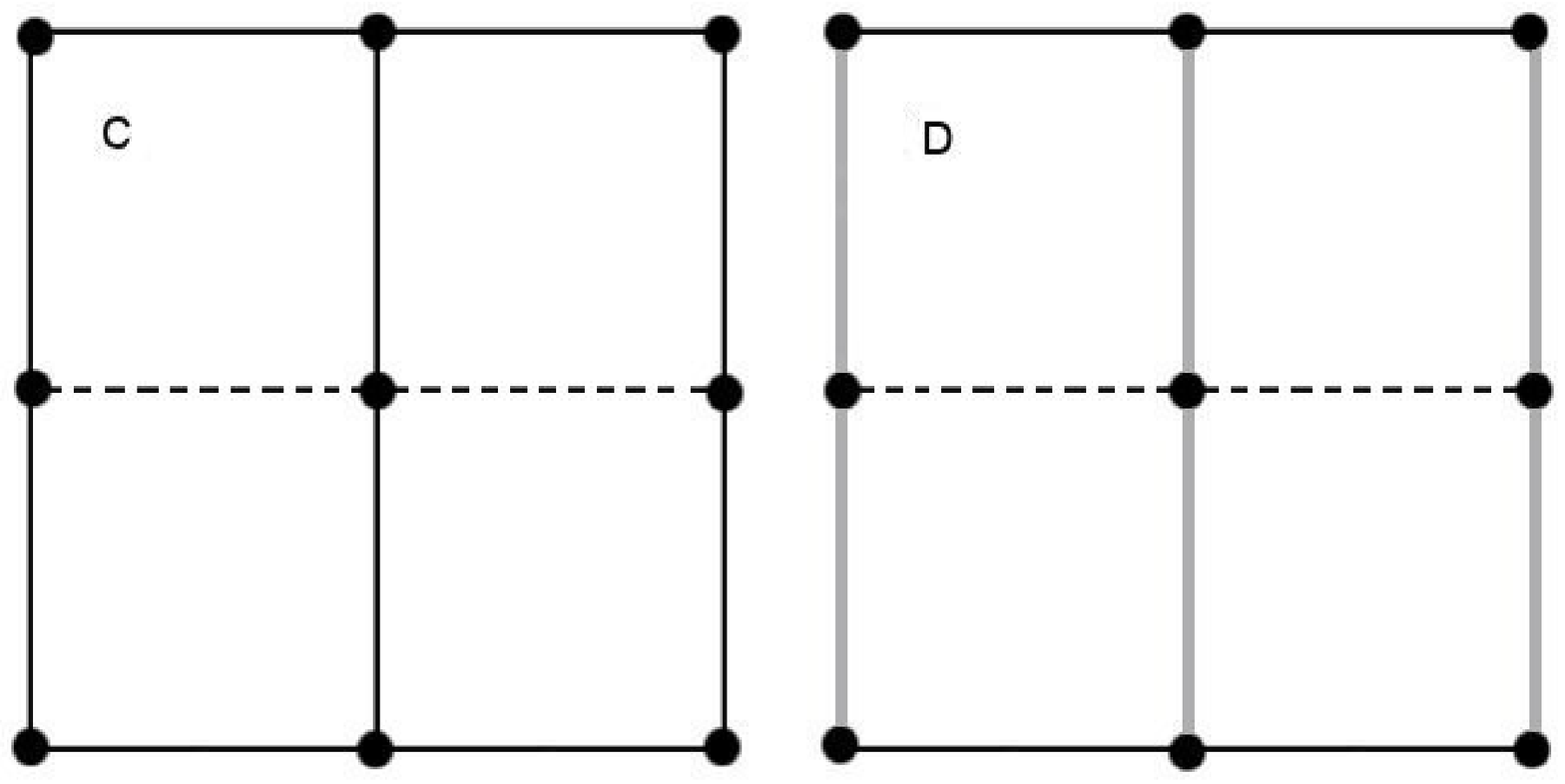}
  \caption{The zero-flux (A and B) and the $\pi$-flux (C and D) mean field
    ans\"atze. The solid black, solid gray and dashed lines indicate bonds
    with $f_{ij}=f$ and $x_{ij}=x$, $f_{ij}=f$ and $x_{ij}=-x$ and $f_{ij}=-f$
    and $x_{ij}=-x$, respectively. Here $f_{ij}=f_{ji}^*$, $x_{ij}=x_{ji}$
    and both $f$ and $x$ are real.}
  \label{possiblestates}
\end{figure}

As discussed before we will consider symmetric spin liquid states that are
invariant under PSG transformations. A PSG transformation will be a
combined transformation of magnetic symmetry group and gauge
transformations. However, it is not possible to combine a magnetic symmetry
transformation with any gauge group transformation since algebraic relations
(\ref{eq:MultiplicationRelations}) between magnetic symmetry group elements
constrain the possible choices of gauge transformations. For each transformation
$T$ $\left(T\in\left\{T_x, T_y, R, \tau \right\}\right)$ in the magnetic
symmetry group there is a gauge group transformation $G_T$
$\left(G_T\in\left\{G_{T_x}, G_{T_y},G_ {R}, G_{\tau} \right\}\right)$ such
that the mean field ansatz is invariant under combined transformation $G_T
T$. The gauge transformations can be represented as follows:
\begin{equation}\label{eq:GaugeTransformationPSG}
G_T: b_{\vec{r}\sigma}\rightarrow e^{i\phi_T\left(\vec{r}\right)}b_{\vec{r}\sigma} .
\end{equation}
The transformations $G_T$ can be found by considering algebraic constraints
(\ref{eq:MultiplicationRelations}) (with $\alpha\rightarrow \alpha_\sigma$
when the magnetic symmetry group transformations are applied to the Schwinger
boson operators $b_{i\sigma}$). For example, since
$e^{i2\pi\alpha_{\sigma}}T_x^{-1}T_yT_xT_y^{-1}=1$, then
$e^{i2\pi\alpha_{\sigma}}\left(G_{T_x}T_x\right)^{-1}G_{T_y}T_yG_{T_x}T_x\left(G_{T_y}T_y\right)^{-1}$
has to be equivalent to an identity operator, that is, it has to be an element of IGG ($1$ or $-1$). Using relation
\begin{equation}\label{eq:PSGEquation1}
Y^{-1}G_XYb_{\vec{r}\sigma}\left(Y^{-1}G_XY\right)^{-1}=e^{i\phi_X\left[Y\left(\vec{r}\right)\right]}b_{\vec{r}\sigma} ,
\end{equation}
where here $X,\;Y \in \left\{T_x, T_y\right\}$, we obtain the following equation
\begin{equation}\label{eq:PSGEquation2}
-\phi_{T_x}\left[T_x\left(\vec{r}\right)\right]+\phi_{T_y}\left[T_x\left(\vec{r}\right)\right]+\phi_{T_x}\left[T_xT_y^{-1}\left(\vec{r}\right)\right]-\phi_{T_y}\left[\vec{r}\right]=p_1\pi ,
\end{equation}
where $p_1 \in \left\{0,\;1\right\}$ corresponds to IGG elements $1$ and $-1$ ($e^{ip_1\pi}$ with $p_1=0$ or $1$).\\
Similar equations can be obtained by considering other algebraic relations in
the equations (\ref{eq:MultiplicationRelations}). A detailed derivation of
those equations is described in Appendix \ref{app:PSG} and the most general
solution that satisfies the obtained equations is:
\begin{eqnarray}\label{eq:PSGSolution}
\phi_{T_x}\left[x,y\right]&=& 0 , \\
\phi_{T_y}\left[x,y\right]&=& p_1\pi x , \nonumber \\
\phi_{\tau}\left[x,y\right]&=& \phi_{\tau}\left[0,0\right]+p_1\pi xy, \nonumber \\
\phi_{R}\left[x,y\right]&=& \phi_R\left[0,0\right]+p_1\pi xy, \nonumber
\end{eqnarray}
where $p_1$ $\in$ $\left\{0, 1\right\}$. If the nearest neighbour amplitudes
$x_{ij}$ are nonzero then
$\phi_R\left[0,0\right]=\phi_{\tau}\left[0,0\right]=p_2\pi/2$ with $p_2$ $\in$
$\left\{0, 1\right\}$. Possible mean field ans\"atze can then be obtained by requiring invariance of the mean field ansatz with respect to PSG transformations:
\begin{eqnarray}\label{eq:MFAnsatz1}
\langle G_T T F_{ij}\left(G_T T\right)^{-1}\rangle=\langle F_{ij}\rangle=f_{ij} , \\
\langle G_T T X_{ij}\left(G_T T\right)^{-1}\rangle=\langle X_{ij}\rangle=x_{ij} , \nonumber
\end{eqnarray}
where $i$ and $j$ are nearest neighbours, $T \in \left\{T_x, T_y, R, \tau
\right\}$ and $f_{ij}$ and $x_{ij}$ are both real and nonzero in general (Appendix \ref{app:PSG}). From the conditions (\ref{eq:MFAnsatz1}) we obtain four possible mean field ans\"atze:
\begin{eqnarray}\label{eq:MFAnsatz2}
  A: \;f_{\left(x_i,y_i\right)\left(x_i+1,y_i\right)}&=&f_{\left(x_i,y_i\right)\left(x_i,y_i+1\right)}=f,\\
 x_{\left(x_i,y_i\right)\left(x_i+1,y_i\right)}&=&x_{\left(x_i,y_i\right)\left(x_i,y_i+1\right)}=x,\nonumber\\
\nonumber\\
B:
\;f_{\left(x_i,y_i\right)\left(x_i+1,y_i\right)}&=&f_{\left(x_i,y_i\right)\left(x_i,y_i+1\right)}=f,\nonumber\\
x_{\left(x_i,y_i\right)\left(x_i+1,y_i\right)}&=&-x_{\left(x_i,y_i\right)\left(x_i,y_i+1\right)}=x,
\nonumber\\
\nonumber\\
C: \; f_{\left(x_i,y_i\right)\left(x_i,y_i+1\right)}&=&\left(-1\right)^{y_i}
f_{\left(x_i,y_i\right)\left(x_i+1,y_i\right)}=f, \nonumber\\
x_{\left(x_i,y_i\right)\left(x_i,y_i+1\right)}&=&\left(-1\right)^{y_i}
x_{\left(x_i,y_i\right)\left(x_i+1,y_i\right)}=x ,\nonumber \\
\nonumber\\
D: \; f_{\left(x_i,y_i\right)\left(x_i,y_i+1\right)}&=&\left(-1\right)^{y_i}
f_{\left(x_i,y_i\right)\left(x_i+1,y_i\right)}=f , \nonumber\\
- x_{\left(x_i,y_i\right)\left(x_i,y_i+1\right)}&=&\left(-1\right)^{y_i} x_{\left(x_i,y_i\right)\left(x_i+1,y_i\right)}=x, \nonumber
\end{eqnarray}
where $f_{ij}=f_{ji}^*$, $x_{ij}=x_{ji}$, $f_{ij},\; x_{ij} \in
\mathbb{R}$ and above equations are valid for all lattice sites
$i=\left(x_i,y_i\right)$. The mean field ans\"atze above, illustrated in Fig. \ref{possiblestates}, then give
local minima of the mean field Hamiltonian. 
The ans\"atze $A$ and $B$ can be distinguished from $C$ and $D$ by the emergent gauge invariant flux through an elementary square plaquette defined as the gauge invariant phase of the following products \cite{TchernyshyovMoessnerSondhi} : 
\begin{eqnarray}\label{eq:Flux}
f_{ij}f_{jk}f_{kl}f_{li}=|f|^4e^{i\Phi} , \\
x_{ij}x_{jk}x_{kl}x_{li}=|x|^4e^{i\Phi} , \nonumber
\end{eqnarray}
with $i$, $j$, $k$, $l$ being sites of a square lattice plaquette. For
the ans\"atze $A$ and $B$ the phase $\Phi=0$ (zero-flux states) and
for $C$ and $D$ the phase $\Phi=\pi$ ($\pi$-flux states) for all
square lattice plaquettes. Therefore these two sets of states are
clearly not gauge equivalent meanfield states. Also, $A$ and $B$ differ
by the symmetry of the Schwinger boson pairing order parameter
$x_{ij}$. Unlike for ansatz $A$, for ansatz $B$ the order parameter
$x_{ij}$ changes sign under a $\pi/2$ rotation ($d$ wave like
state). For $\pi$-flux states, $C$ and $D$, the unit cell for
Schwinger bosons is doubled due to the emergent $\pi$ flux through
each unit square plaquette of the lattice and the translation symmetry
is not explicit in the mean field Hamiltonian. Similarly as for the
ans\"atze $A$ and $B$, ans\"atze $C$ and $D$ differ by the symmetry of
the order parameter $x_{ij}$.
In the following section we further study possible noncondensed phases arising from the ans\"atze described previously and relate these states to bosonic fractional quantum Hall states for bosonic atoms in rotating optical lattices.\cite{MollerCooper} 

\section{Bosonic fractional quantum Hall states}
\label{sec:FQH}
To further study noncondensed phases of the mean field Hamiltonian we take into account the local constraint (\ref{eq:Constraint}) on Schwinger boson number at each site exactly,\cite{ZouDoucotShastry, HallbergRojoBalseiro, KrivnovLikhachevOvchinnikov, ChangYang} unlike in the conventional Schwinger boson mean field theory where the constraint is imposed on average as in the equation (\ref{eq:ConstarintAverage}). The local constraint (\ref{eq:Constraint}) introduces strong correlations between spin-up and spin-down bosons. First we note that the constraint implies that the number of up/down bosons at each site can be only zero or one. Thus we require that the Schwinger boson operators satisfy the following anticommutation relations on the same site:
\begin{equation}\label{eq:AnticommutationRelations}
\left\{a_{i,\sigma}, a_{i,\sigma'}\right\}_+=0 \;,\; \left\{a_{i,\sigma}, a_{i,\sigma'}^\dagger\right\}_+=\delta_{\sigma,\sigma'},
\end{equation}
where $\sigma,\sigma'\in \left\{ \uparrow,\downarrow \right\}$, and commutation relations 
\begin{equation}\label{eq:CommutationRelations}
\left[a_{i,\sigma}, a_{j,\sigma'}\right]_-=0 \;,\; \left[a_{i,\sigma}, a_{j,\sigma'}^\dagger\right]_-=0,
\end{equation}
for the lattice sites $i$ and $j$ when $i\ne j$. The requirement above corresponds to a hard-core limit for each bosonic species and excludes states with more than one boson of the same species at the same lattice site. However, to take the constraint (\ref{eq:Constraint}) into account exactly we also need to exclude the states with one spin-up and one spin-down Schwinger boson at the same lattice site. This is achieved by the Gutzwiller projection:
\begin{equation}\label{eq:GutzwillerProjection}
P_{G}=\prod_i(1-n_{i\uparrow} n_{i\downarrow}),
\end{equation}
where $n_{i\uparrow}$ and $n_{i\downarrow}$ are the number operators at a site $i$ for spin-up and spin-down Schwinger bosons respectively, which for hard-core bosons have eigenvalues $0$ or $1$. The ground state wave function is then 
\begin{equation}\label{eq:Wavefunction}
\psi=P_G \psi_{MF} ,
\end{equation} 
where $\psi_{MF}$ is the mean field Hamiltonian ground state wave
function. To apply the Gutzwiller projection explicitly it is 
convenient to rewrite the mean field Hamiltonian
(\ref{eq:Hamiltonian_MF}) in the following form:
\begin{widetext}
  \begin{eqnarray}\label{eq:MFHamiltonian2}
    H_{MF}&=&h_0+\frac{h}{2}\sum_i\left(p_{i\downarrow}^{\dagger}p_{i\downarrow}+h_{i\uparrow}^{\dagger}h_{i\uparrow}\right)-\frac{J}{4}\sum_{\langle
      ij\rangle}
    f_{ij}\left\{e^{i\phi_{ij}/2}\left(p_{i\downarrow}p_{j\downarrow}^{\dagger}+h_{i\uparrow}h_{j\uparrow}^{\dagger}\right)+e^{-\phi_{ij}/2}\left(p_{j\downarrow}p_{i\downarrow}^{\dagger}+h_{j\uparrow}h_{i\uparrow}^{\dagger}\right)\right\}
    \nonumber\\
    &&-\frac{J}{4}\sum_{ \langle ij \rangle} x_{ij}\left\{e^{i\phi_{ij}/2}\left(p_{i\downarrow}h_{j\uparrow}^{\dagger}+h_{i\uparrow}p_{j\downarrow}^{\dagger}\right)+e^{-i\phi_{ij}/2}\left(h_{j\uparrow}p_{i\downarrow}^{\dagger}+p_{j\downarrow}h_{i\uparrow}^{\dagger}\right)\right\},\nonumber
  \end{eqnarray}
\end{widetext}
where 
\begin{equation}\label{eq:HConstant}
h_0=\frac{J}{4}\sum_{\langle
  ij\rangle}\left(|f_{ij}|^2+|x_{ij}|^2\right)+\frac{J}{4}\sum_{\langle
  ij\rangle}n_{i}\cdot n_j+\frac{Jz}{8}\sum_i n_i -N_s\frac{h}{2},
\end{equation}
with $N_s$ being the number of lattice sites, and where we have introduced the notation $p_{i\downarrow}=a_{i\downarrow}$ and $h_{i\uparrow}=a_{i\uparrow}^{\dagger}$. We have also taken into account that $f_{ij}$ and $x_{ij}$ are real for all mean field ans\"atze (\ref{eq:MFAnsatz2}). The operator $h_{i\uparrow}^{\dagger}$ creates a spin-up bosonic hole and the operator $p_{i\downarrow}^{\dagger}$ creates a spin-down bosonic particle at the site $i$. The constraint (\ref{eq:Constraint}) can then be rewritten in the form:
\begin{equation}\label{eq:Constraint2}
p_{i\downarrow}^{\dagger}p_{i\downarrow}=h_{i\uparrow}^{\dagger}h_{i\uparrow},
\end{equation}
where we have used the anticommutation relations
(\ref{eq:AnticommutationRelations}). The local constraint
(\ref{eq:Constraint2}) is equivalent to saying that wherever there is a
spin-down boson there is also a spin-up hole. The Gutzwiller projected
wave function with $N_{\downarrow}$ spin-down bosons and
$N-N_{\downarrow}$ spin-up bosons can then be obtained in the
following way. Starting from the state with $N$ spin-up bosons, first
$N_{\downarrow}$ spin-up bosons are annihilated and then
$N_{\downarrow}$ spin-down bosons created at the coordinates of the
spin-up holes. Therefore the wave function in the coordinate
representation will be a function of the coordinates of spin-down
bosons (or equivalently of spin-up bosonic holes) only.

Before examining further the effect of the Gutzwiller projection we
note that the Hamiltonian (\ref{eq:MFHamiltonian2}) describes
correlated hopping of particle-hole pairs. We are thus led to introduce the following hard core boson operators:
\begin{equation}\label{eq:PlusMinusOperators}
b_{i\pm}=\frac{1}{\sqrt{2}}\left(p_{i\downarrow}\pm h_{i\uparrow}\right).
\end{equation}
It can be easily verified that these operators satisfy mixed commutation - anticommutation relations of the form (\ref{eq:AnticommutationRelations}) and (\ref{eq:CommutationRelations}). In terms of operators (\ref{eq:PlusMinusOperators}) the mean field Hamiltonian (\ref{eq:MFHamiltonian2}) can be written as:
\begin{eqnarray}\label{eq:MFHamiltonian3}
&&H_{MF}=h_0+\frac{h}{2}\sum_i\left(b_{i+}^{\dagger}b_{i+}+b_{i-}^{\dagger}b_{i-}\right) \\
&-&\frac{J}{4}\sum_{\langle
  ij\rangle}
  t_{ij}^{+}\left\{e^{i\phi_{ij}/2}b_{i+}b_{j+}^{\dagger}+e^{-i\phi_{ij}/2}b_{j+}b_{i+}^{\dagger}\right\} \nonumber\\
&-&\frac{J}{4}\sum_{\langle
  ij\rangle}
  t_{ij}^{-}\left\{e^{i\phi_{ij}/2}b_{i-}b_{j-}^{\dagger}+e^{-i\phi_{ij}/2}b_{j-}b_{i-}^{\dagger}\right\}, \nonumber
\end{eqnarray}
where $t_{ij}^{\pm}=f_{ij}\pm x_{ij}$. Also, the constraint (\ref{eq:Constraint2}) implies the following condition:
\begin{equation}\label{eq:Constraint3}
b_{i+}^{\dagger}b_{i+}=b_{i-}^{\dagger}b_{i-}.
\end{equation}
This can be seen from the equation
$b_{i+}^{\dagger}b_{i+}-b_{i-}^{\dagger}b_{i-}=p_{i\downarrow}^{\dagger}h_{i\uparrow}+h_{i\uparrow}^{\dagger}p_{i\downarrow}$. When
the constraint (\ref{eq:Constraint2}) is taken into account explicitly then
the only allowed states are the states with zero or one particle-hole pairs at
a lattice site. Within this restricted single site basis terms
$p_{i\downarrow}^{\dagger}h_{i\uparrow}$ and
$h_{i\uparrow}^{\dagger}p_{i\downarrow}$ can be set to zero and condition
(\ref{eq:Constraint3}) is obtained.\\
The ground state wave function can then be obtained through the following
projection:
\begin{equation}\label{eq:Wavefunction2}
\psi\left(\left\{\vec{r}_i\right\}\right)=\langle 0|\prod_{i=1}^N
h_{i\uparrow}p_{i\downarrow}|MF\rangle=\langle 0|\prod_{i=1}^{N}b_{i+}b_{i-}|MF\rangle,
\end{equation}
where $|0\rangle$ represents the vacuum of spin-down bosons and spin-up holes,
(or equivalently vacuum of $b_{+}$ and $b_{-}$ bosons), $\vec{r}_i$ are
coordinates of $N$ bosonic atoms in the lattice, and $|MF\rangle$ is a mean
field state of bosons $b_{+}$ and $b_{-}$. \\
The effective flux per plaquette of the lattice for $b_{+}/b_{-}$ bosons is
$\alpha_{eff}^{\pm}=\alpha_{eff}/2$, with $\alpha_{eff}$ being the effective
flux for the atoms. For the mean field ans\"atze $A$ and $B$ the effective
flux $\alpha_{eff}^{\pm}=\alpha/2$, and  for the mean field anz\"atze $C$
and $D$ $\alpha_{eff}^{\pm}=\alpha/2+1/2$. Since the number of atoms is the
same as the number of $b_{+}$ ($b_{-}$) bosons and
$\alpha_{eff}^{\pm}=\alpha_{eff}/2$ the wavefunction can be written as:
\begin{equation}\label{eq:Wavefunction3}
\psi_{\nu}\left(\left\{\vec{r}_i\right\}\right)=\left[\psi'_{\nu'=2\nu}\left(\left\{\vec{r}_i\right\}\right)\right]^2,
\end{equation}
where $\psi'_{\nu'}\left(\left\{\vec{r}_i\right\}\right)$ is the wavefunction
for $b_{+}$/$b_{-}$ bosons at filling fractions
$\nu'=n_+/\alpha_{eff}^+=n_{-}/\alpha_{eff}^{-}=2\nu=2n/\alpha_{eff}$, with
$n_+$/$n_{-}$ being the average density of $b_{+}$/$b_{-}$ bosons per lattice
site. Thus at an atom filling fraction $\nu$, the average density of atoms per
lattice site is $n=\alpha\nu'/2$ for the mean field ans\"atze $A$ and $B$, and
$n=\left(1/2\pm\alpha/2\right)\nu'$ for the mean field ans\"atze $C$ and
$D$, where we have taken into account the symmetries under
$\alpha_{eff}^{\pm}\rightarrow 1-\alpha_{eff}^{\pm}$ and $n\rightarrow 1-n$
(particle-hole symmetry on the lattice). \\
We further argue that the FQH states of bosonic atoms in a rotating optical
lattice correspond to the incompressible states for $b_{+}-b_{-}$
pairs. Without the constraint, two species of hard-core bosons, $b_+$ and
$b_-$, are independent of each other and the ground state wavefunction of the
mean-field Hamiltonian (\ref{eq:MFHamiltonian3}) can be written as a product
of the ground state wavefunctions for $b_+$ and $b_-$ bosons. For example, at
an effective filling fraction $\nu_{eff}'=1$ the ground state for each bosonic
species ($b_+$ and $b_-$) is predicted to be an incompressible state that, in
the continuum limit, can be described by the bosonic Pfaffian state.\cite{WilkinGunn} The
projected wavefunction (\ref{eq:Wavefunction3}) then corresponds to the
$\nu_{eff}=1/2$ FQH states for $b_{+}-b_{-}$ pairs
(or equivalently for particle-hole pairs). 
\\
At $\nu_{eff}'=1$, that is $\nu_{eff}=1/2$, possible incompressible ground states are at $n=\alpha/2$
(mean field ans\"atze $A$ and $B$) and $n=1/2\pm\alpha/2$ (mean field
ans\"atze $C$ and $D$). The states with $n=1/2\pm\alpha/2$ are lattice induced
FQH states, with no counterpart in the continuum, and within presented mean
field theory, correspond to ``$\pi$
flux'' spin liquid states of the effective spin model (mean field ans\"atze
$C$ and $D$) for which the hard-core bosons $b_+$ and $b_-$ see an effective
flux $\alpha_{eff}^\pm=1/2+\alpha/2$. Such incompressible quantum Hall states have
previously been found using the composite fermion approach and exact
diagonalization studies.\cite{MollerCooper} Several other lattice induced
incompressible states predicted within the composite fermion approach can also
be related to incompressible states for $b_+-b_-$ pairs, for example states at
$\nu_{eff}'=2/3$.\\
Finally, we note that the projected wavefunctions are also classified by PSG
and that our results provide a guideline for a further search of possible ground
states in the projected wavefunction space. We thus propose to study the
effective spin model by Gutzwiller projected wavefunction variational approach
\cite{Gros,Liu,SensarmaGalitski} as a possible direction of future research. 

\section{Conclusions}
\label{sec:Conclusions}
We have studied possible incompressible ground states for
bosonic atoms in a rotating square optical lattice. In the hard-core boson
limit the Bose-Hubbard Hamiltonian for the system can be mapped to a
frustrated easy-plane magnet. Using Schwinger boson mean field theory and
projective symmetry group analysis we have classified possible stable gapped
spin liquid states of the effective spin model and related those states to FQH
states for bosonic atoms in rotating optical lattices. In particular, we have
found that, within the mean field theory developed here, the FQH states induced by
lattice potential and with no counterpart in the continuum limit, correspond
to ``$\pi$ flux'' spin liquid states of the effective spin model. Since there
are also competing condensed states on the lattice
\cite{DuricLee,Palmer,Kasamatsu} it is important to test the
predictions of the mean field theory presented here. Finally, further work is
necessary for more direct comparison of the obtained results with known
results for FQH states in rotating optical lattices. We hope the Gutzwiller
projected wavefunction variational approach will be further developed in future
research. 

\section{Acknowledgments}
\label{sec:acknowledgments}
We thank Pedro Ribeiro for carefully reading the manuscript and for very
helpful discussions. We also thank Paul McClarty and Dmitry Kovrizhin for very 
helpful discussions and suggestions.  
\appendix

\section{Solution to PSG equations}
\label{app:PSG}
In this appendix we derive equations for allowed gauge transformations $G_T$,
defined by equation (\ref{eq:GaugeTransformationPSG}) in section \ref{sec:PSG}:
\begin{equation}\label{eq:A1}
G_T: b_{\vec{r}\sigma}\rightarrow e^{i\phi_T\left(\vec{r}\right)}b_{\vec{r}\sigma} .
\end{equation}
As explained in section \ref{sec:PSG} the mean field ansatz should be
invariant under combined transformations $G_TT$ where $T$ are magnetic
symmetry group transformations. The transformations $G_T$ can be found by
considering algebraic constraints (\ref{eq:MultiplicationRelations}) following
the procedure explained in the example in the main text. We first describe in
more details derivation of the equation ( \ref{eq:PSGEquation2}) and then
apply the same procedure to all other algebraic constraints given by equations
(\ref{eq:MultiplicationRelations}).\\
From 
\begin{equation}\label{eq:A2}
e^{i2\pi\alpha_{\sigma}}T_x^{-1}T_yT_xT_y^{-1}=1 
\end{equation}
it follows
\begin{equation}\label{eq:A3}
e^{i2\pi\alpha_{\sigma}}(G_{T_x}T_x)^{-1}G_{T_y}T_yG_{T_x}T_x(G_{T_y}T_y)^{-1}=e^{ip_1\pi}
\end{equation}
where $p_1\in \left\{ 0,1 \right\}$ corresponds to IGG elements $1$ and
$-1$. The relation (\ref{eq:A3}) can be rewritten as 
\begin{eqnarray}\label{eq:A4}
&&T_x^{-1}G_{T_x}^{-1}T_xT_x^{-1}G_{T_y}T_xT_x^{-1}T_yT_xT_y^{-1}T_yT_x^{-1}G_{T_x}T_xT_y^{-1}G_{T_y}^{-1}\nonumber\\
&&=e^{ip_1\pi}e^{-i2\pi\alpha_{\sigma}}. 
\end{eqnarray}
Since $T_xT_x^{-1}=1$, and $T_x^{-1}T_yT_xT_y^{-1}=e^{-i2\pi\alpha_{\sigma}}$
and 
\begin{equation}\label{eq:A5}
Y^{-1}G_xYb_{\vec{r}\sigma}\left(Y^{-1}G_xY\right)^{-1}=e^{i\phi_X\left[Y\left(\vec{r}\right)\right]}b_{\vec{r}\sigma}
\end{equation}
we obtain equation (\ref{eq:PSGEquation2}):
\begin{equation}\label{eq:A6}
-\phi_{T_x}\left[T_x\left(\vec{r}\right)\right]+\phi_{T_y}\left[T_x\left(\vec{r}\right)\right]+\phi_{T_x}\left[T_xT_y^{-1}\left(\vec{r}\right)\right]-\phi_{T_y}\left[\vec{r}\right]=p_1\pi ,
\end{equation}
with $p_1\in\left\{0,1\right\}$. We also note that, since $\phi_T$ are phases,
all equations for phases in this section will be true modulo $2\pi$. If a gauge transformation
$b_{\vec{r}\sigma}\rightarrow e^{i\phi_G\left(\vec{r}\right)}
b_{\vec{r}\sigma}$ is applied then phases in (\ref{eq:A1}) change as follows:
\begin{equation}\label{eq:A7}
\phi_T\left(\vec{r}\right)\rightarrow\phi_T\left(\vec{r}\right)+\phi_G\left(\vec{r}\right)-\phi_G\left[T^{-1}\left(\vec{r}\right)\right]. 
\end{equation}
Using this gauge freedom the mean field ansatz can be made explicitly
invariant under translation $T_x$ in $x$ direction, that is we can assume 
\begin{equation}\label{eq:A8}
\phi_{T_x}\left(\vec{r}\right)=0. 
\end{equation}
Then the equation (\ref{eq:A6}) becomes:
\begin{equation}\label{eq:A9}
\phi_{T_y}\left[x+1,y\right]-\phi_y\left[x,y\right]=p_1\pi. 
\end{equation} 
The solution of the equation above is then:
\begin{equation}\label{eq:A10}
\phi_{T_y}\left[x,y\right]=\phi_{T_y}\left[0,y\right]+p_1\pi x.
\end{equation}
We further assume that using gauge freedom we can set
$\phi_{T_y}\left[0,y\right]=0$ while keeping
$\phi_{T_1}\left(\vec{r}\right)=0$. These two equations can be simultaneously
satisfied for open boundary condition. We further assume open boundary
condition in which case equation (\ref{eq:A10}) becomes:
\begin{equation}\label{eq:A11}
\phi_{T_y}\left[x,y\right]=p_1\pi x.
\end{equation}\\
Following the same procedure, from the algebraic constraints
$T_x^{-1}RT_{y}^{-1}R^{-1}=1$ and $T_{y}^{-1}RT_xR^{-1}=1$ we obtain:
\begin{equation}\label{eq:A12}
\phi_R\left[x+1,y\right]-\phi_R\left[x,y\right]=p_2'\pi+p_1\pi y,
\end{equation}
and 
\begin{equation}\label{eq:A13}
\phi_R\left[x,y+1\right]-\phi_R\left[x,y\right]=p_3'\pi+p_1\pi x,
\end{equation}
with $p_2'$, $p_3'$ $\in$ $\left\{0,1\right\}$.
The solution of the above equations is:
\begin{equation}\label{eq:A14}
\phi_R\left[x,y\right]=\phi_R\left[0,0\right]+p_2'\pi x +p_3'\pi y+ p_1\pi xy, 
\end{equation}
Equivalently from $\tau T_y=T_x\tau$ and $\tau T_x=T_y\tau$ (or
$\tilde{\tau}T_y=T_x\tilde{\tau}$ and
$e^{i4\pi\alpha_{\sigma}y}\tilde{\tau}T_x=T_y\tilde{\tau}$) we obtain
following equations:
\begin{equation}\label{eq:A15}
\phi_{\tau}\left[x+1,y\right]-\phi_{\tau}\left[x,y\right]=p_4'\pi+p_1\pi y
\end{equation}
and 
\begin{equation}\label{eq:A16}
\phi_{\tau}\left[x,y+1\right]-\phi_{\tau}\left[x,y\right]=p_5'\pi+p_1\pi x,
\end{equation}
with $p_4'$, $p_5'$ $\in$ $\left\{0,1\right\}$.
The solution of the equations (\ref{eq:A15}) and (\ref{eq:A16}) is:
\begin{equation}\label{eq:A17}
\phi_{\tau}\left[x,y\right]=\phi_{\tau}\left[0,0\right]+p_4'\pi x+p_5'\pi
y+p_1\pi xy. 
\end{equation}
If we further apply the gauge transformation $\phi_{G}\left[x,y\right]=\pi x$
the phases $\phi_{T}$ will change as follows:
\begin{eqnarray}\label{eq:A18}
\phi_{T_x}\left[x,y\right]&\rightarrow& \phi_{T_x}\left[x,y\right]+\pi=\pi,\\
\phi_{T_y}\left[x,y\right]&\rightarrow&\phi_{T_y}\left[x,y\right]=p_1\pi x,\nonumber\\
\phi_{R}\left[x,y\right]&\rightarrow&
\phi_R\left[x,y\right]+\pi\left(x-y\right)=\phi_R\left[0,0\right]+\left(p_2'+1\right)\pi
x \nonumber\\
&+&\left(p_3'-1\right)\pi y+p_1\pi xy,\nonumber\\
\phi_{\tau}\left[x,y\right]&\rightarrow&\phi_{\tau}\left[x,y\right]+\pi\left(x-y\right)=\phi_R\left[0,0\right]+\left(p_4'+1\right)\pi
x \nonumber\\
&+&\left(p_5'-1\right)\pi y+p_1\pi xy.\nonumber
\end{eqnarray}  
This gauge transformation does not change $\phi_{T_y}\left(\vec{r}\right)$,
nor does it really change $\phi_{T_x}\left(\vec{r}\right)$ since a site-independent
constant $\pi$ can be added to any $\phi_T\left(\vec{r}\right)$ because of the
IGG structure. However, when the transformation is applied
$p_{2}'$, $p_{4}'$  $\rightarrow$ $ p_2'+1$, $p_4'+1$, and $p_3'$, $p_5'$
$\rightarrow$ $p_3'-1$, $p_5'-1$. Since $p'_i=0$ and
$p'_i= 1$ are gauge equivalent $\forall$ $i\in\left\{2,3,4,5\right\}$ we can
  always assume $p_2'=p_3'=p_4'=p_5'=0$.\\
Then we can write:
\begin{eqnarray}\label{eq:A19}
\phi_{T_x}\left[x,y\right]&=&0,\\
\phi_{T_y}\left[x,y\right]&=&p_1\pi x,\nonumber\\
\phi_{R}\left[x,y\right]&=&\phi_{R}\left[0,0\right]+p_1\pi xy, \nonumber\\
\phi_{\tau}\left[x,y\right]&=&\phi_{\tau}\left[0,0\right]+p_1\pi xy.\nonumber
\end{eqnarray}
The allowed mean field ans\"atze can then be obtained by requireing invariance of
the mean field ansatz with respect to PSG transformations:
\begin{eqnarray}\label{eq:A20}
\langle G_T T F_{ij}\left(G_T T\right)^{-1}\rangle=\langle F_{ij}\rangle=f_{ij} , \\
\langle G_T T X_{ij}\left(G_T T\right)^{-1}\rangle=\langle X_{ij}\rangle=x_{ij} , \nonumber
\end{eqnarray}
where $i$ and $j$ are nearest neighbours and $T \in \left\{T_x, T_y, R, \tau
\right\}$.\\ Considering further following equations for the 
$\left(0,0\right)\rightarrow\left(-1,0\right)$ bond:
$f_{\left(0,0\right)\left(-1,0\right)}=e^{-i\left(\phi_R\left[0,1\right]+\phi_R\left[1,0\right]-2\phi_R\left[0,0\right]\right)}f_{\left(0,0\right)\left(1,0\right)}=f_{\left(0,0\right)\left(1,0\right)}$
and
$f_{\left(-1,0\right)\left(0,0\right)}=f_{\left(0,0\right)\left(1,0\right)}=f^*_{\left(0,0\right)\left(-1,0\right)}$
we see that
$f_{\left(0,0\right)\left(-1,0\right)}=f^*_{\left(0,0\right)\left(-1,0\right)}\equiv
f$. In other words $f_{\left(0,0\right)\left(-1,0\right)}$ has to be
real. Also, a global phase rotation, that is, a gauge transformation
$\phi_G\left(\vec{r}\right)=\phi$ does not change any PSG elements (equation
(\ref{eq:A7})). This can be used to fix one of the $x_{ij}$ presented to be real and
positive, for example,
$x_{\left(0,0\right)\left(0,1\right)}=x^*_{\left(0,0\right)\left(0,1\right)}\equiv
x$.\\
If we assume that the nearest neighbour amplitudes $x_{ij}$ are nonzero then
$\phi_{R}\left[0,0\right]$ can be found from the following equation for the
$\left(0,0\right)\rightarrow\left(-1,0\right)$ bond: $x_{\left(0,0\right) \left(-1,0\right)}=e^{-i\left(2\phi_R\left[0,0\right]+\phi_R\left[1,0\right]+\phi_R\left[0,1\right]\right)}x_{\left(0,0\right)\left(1,0\right)}$,
which in combination with the equations
$x_{\left(0,0\right)\left(-1,0\right)}=x_{\left(-1,0\right)\left(0,0\right)}=x_{\left(0,0\right)\left(1,0\right)}$
(due to the symmetry $x_{ij}=x_{ji}$ and translational invariance of $x_{ij}$ in
$x$ direction) gives $4\phi_R\left[0,0\right]=0$, that is,
$\phi_R\left[0,0\right]=p_{2}\pi/2$ with $p_2$ $\in$
$\left\{0,1\right\}$. Equivalently from
$x_{\left(0,0\right)\left(0,1\right)}=e^{-i\left(\phi_R\left[0,0\right]+\phi_R\left[1,0\right]\right)}x_{\left(0,0\right)\left(1,0\right)}=e^{-i\left(\phi_\tau\left[0,0\right]+\phi_\tau\left[-1,0\right]\right)}x_{\left(0,0\right)\left(-1,0\right)}$
we obtain $\phi_{\tau}\left[0,0\right]=\phi_R\left[0,0\right]$. The algebraic
constraints $R^4=1$, $R\tau R\tau=1$ and $\tau\tau=1$ do not further restrict
possible values of $p_1$ and $p_2$. The mean field ans\"atze
(\ref{eq:MFAnsatz1}) can then be obtained from equations (\ref{eq:A19}) and
(\ref{eq:A20}) with $\phi_R\left[0,0\right]=\phi_{\tau}\left[0,0\right]=p_2
\pi/2$.

\end{document}